# Public preferences for marine plastic litter reductions across Europe

Salma Khedr,[1*] Katrin Rehdanz,[1] Roy Brouwer,[2,3] Hanna Dijkstra,[3] Sem Duijndam[3], Pieter van Beukering[3], and Ikechukwu C. Okoli[1]

[1] Kiel University, Department of Economics, Wilhelm-Seelig-Platz 1, 24118 Kiel, Germany

[2] Department of Economics and the Water Institute, University of Waterloo, 200 University Avenue West, Waterloo, Ontario, N2L 3G1, Canada

[3] Institute for Environmental Studies (IVM), Vrije Universiteit Amsterdam, De Boelelaan 1111, 1081 HV Amsterdam, The Netherlands

*Corresponding author: khedr@economics.uni-kiel.de

**Abstract**

Plastic pollution is one of the most challenging problems affecting the marine environment of our time. Based on a unique dataset covering four European seas and eight European countries, this paper adds to the limited empirical evidence base related to the societal welfare effects of marine litter management. We use a discrete choice experiment to elicit public willingness-to-pay (WTP) for macro and micro plastic removal to achieve Good Environmental Status across European seas as required by the European Marine Strategy Framework Directive. Using a common valuation design and following best-practice guidelines, we draw meaningful comparisons between countries, seas and policy contexts. European citizens have strong preferences to improve the environmental status of the marine environment by removing both micro and macro plastic litter favouring a pan-European approach. However, public WTP estimates differ significantly across European countries and seas. We explain why and discuss implications for policymaking.

**Acknowledgements**

This work was funded under the EU Horizon 2020 project CLAIM: Cleaning Litter by developing and Applying Innovative Methods in European seas (Grant agreement number 774586). We are grateful to our colleagues in the CLAIM project who provided feedback on early versions of the online questionnaire, and the colleagues and students who helped us translate the questionnaire in different languages.

# 1 Introduction

Plastic pollution is omnipresent in the global marine environment (Jambeck et al., 2015; Jeftic et al., 2009), and has been documented to negatively affect organisms, ecosystems, and human wellbeing (Gregory, 2009; Rochman et al., 2013; Troost et al., 2018; Werner et al., 2016), highlighting the need for policy action (GESAMP, 2015; UNEP, 2016). In the European Union (EU), the Marine Strategy Framework Directive (MSFD) is the main legal instrument to protect the EU's marine environment (Frantzi et al., 2021). Marine litter is one of the eleven qualitative descriptors listed in Annex I of the MSFD that the Directive targets to achieve Good Environmental Status (GES). GES has to be achieved by 2020 and is defined in terms of marine litter as "properties and quantities of marine litter [that] do not cause harm to the coastal and marine environment" (European Commission Directive 2008/56/EC, Annex I). While marine litter refers to any persistent, manufactured or processed solid material discarded, disposed of or abandoned in the marine and coastal environment, in the EU's marine environment this litter mostly consists of plastics. For example, in the Baltic Sea 70% of marine litter is plastic (HELCOM, 2018). Similarly, 82% of the marine litter found in the Mediterranean Sea was found to be comprised of plastic material (Vlachogianni, 2019).

The MSFD requires EU Member States to develop and implement strategies focusing on so-called programmes of measures to attain or preserve the marine environment's GES. Further, Member States have to evaluate these programmes of measures by examining their cost effectiveness, and justify their economic rationality by carrying out cost-benefit analyses (Bertram et al., 2014; Börger et al., 2016). The costs of measures are usually relatively easy to calculate, however, estimating the associated benefits, which are necessary to conduct a cost-benefit analysis, can prove challenging. This is due to the public nature of many of the ecosystem goods and services provided under GES, requiring monetary valuation of non-market benefits, such as the cultural ecosystem services related to the existence of habitats for species, improvements in recreational opportunities, and aesthetic values (Nieminen et al., 2019). Moreover, challenges arise due to the indirect nature of some benefits, for instance a change in the overall ecosystem health of the marine environment, which will eventually lead to a change in the ecosystem's ability to provide goods and services to beneficiaries, such as fisheries and tourism. Therefore, the overall change in social welfare due to changes in the marine environment may be difficult to assess (Bertram et al., 2014).

Various studies provide information on the broader societal benefits of programmes of measures taken to achieve GES in the marine environment. However, these studies mostly focus on eutrophication (e.g., Ahtiainen and Vanhatalo, 2012; Ahtiainen et al., 2015), another descriptor of GES in the MSFD. Studies focusing on the economic valuation of the societal benefits of marine litter reduction are few, as a recent review of the literature demonstrates (Stöver et al., 2021). In their review of 22 studies, Stöver and colleagues identified 6 non-market (stated preferences) valuation studies that provide value estimates for 7 of the 27 EU Member States. Two of these six studies focus on beach litter (Blakemore et al., 2002; Brouwer et al., 2017), while two studies cover coastal waters or the marine environment more generally (Hynes et al., 2013; Latinopolous et al., 2018). Further evidence is provided in two studies that focus on archipelago areas (Österberg et al., 2012, 2013). Regarding the type of litter, only three of the six studies explicitly mention plastic litter (Brouwer et al., 2017; Hynes et al., 2013; Latinopolous et al., 2018).

The main objective of this study is to estimate the public benefits of marine litter management policies across different European seas by deploying a stated preference discrete choice experiment (DCE). We add to the scarce literature and empirical evidence base on the benefits of marine litter reduction by focusing on plastic litter and distinguishing between micro and macro plastic in a large-scale multi-country DCE. To our knowledge, this is the first study to explicitly assess and compare public preferences and valuation of reductions of micro plastics and macro plastics. This is also the first study covering four major regional seas in Europe in one and the same survey by sampling users and non-users of coastal areas from eight different EU countries (Estonia, Denmark, Germany and Sweden bordering the Baltic Sea; Germany and The Netherlands situated along the North Sea; France located along the Atlantic Ocean, and Greece, France and Italy bordering the Mediterranean Sea). The uniqueness of our dataset in terms of its focus on micro and macro plastics, geographical scope and sample characteristics allows for a comprehensive comparison across countries and contexts. Finally, the design of the DCE enables us to determine public willingness-to-pay (WTP) for achieving GES for these European seas in terms of marine litter reduction in line with the goals of the MSFD for the individual seas separately and for all European seas together.

The remainder of the paper is structured as follows. Section 2 provides an overview of our study area. Section 3 presents the general survey design and a description of the econometric models. Section 4 presents the survey results and the scenario analysis. Finally, Section 5 puts our results in perspective of the literature and concludes.

## 2 Marine litter in European seas

In this section, we provide an overview of the available information related to marine litter present in the European seas investigated in this study (see Figure 1). Although the MSFD requires that EU Member States monitor their progress towards achieving GES, not all indicators are monitored systematically, especially not marine litter. Contrary to marine litter, beach litter has been monitored in some European countries for years (Addamo et al., 2017). However, the first comprehensive overview of beach litter, including the amount of litter, litter categories and trends at different spatial scales from local to EU level, was only recently published using 2015/2016 as baseline years (Hanke et al., 2019). This baseline dataset includes macro litter and litter fragments of at least 2.5 cm. Beyond beach litter, the development of baselines for other types of litter is ongoing in different Member States, including the development of assessment methods for seafloor litter micro litter.[1] In particular, information on micro plastics is often only available based on incidental studies because no monitoring infrastructure exists yet for micro litter (Tekman et al., 2021).

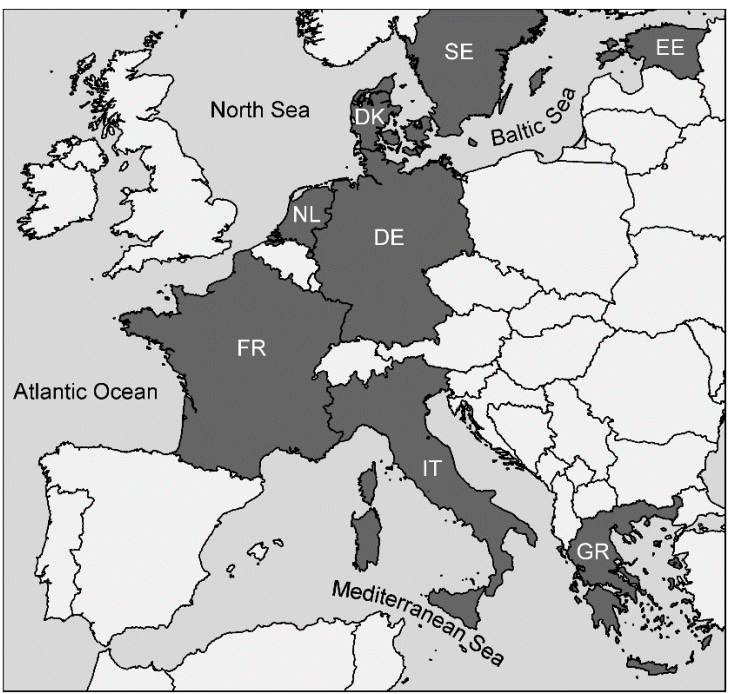

*Figure 1: Overview of the 8 European countries surveyed in this study and the 4 regional European seas bordered by these countries*

In view of the fact that regional seas are transboundary, four Regional Sea Conventions were established in Europe several decades ago to jointly monitor and manage these seas. The work presented here covers the jurisdiction of three of the four conventions: the OSPAR Convention for the Protection of the Marine Environment in the North-East Atlantic, the Helsinki Convention (HELCOM) on the Protection of the

---

[1] See https://mcc.jrc.ec.europa.eu/main/dev.py?N=41&O=434&titre_chap=TG%20Litter (last retrieved 27 April, 2021).

Marine Environment in the Baltic Sea Area, and the Barcelona Convention for the Protection of Marine Environment and the Coastal Region of the Mediterranean. These conventions play a key role in the implementation of the MSFD, which requires that Member States make use of the available institutional structures and activities of the conventions, while the European Commission consults with the conventions to develop standards and criteria on GES.

OSPAR divides the North East Atlantic into five regions, namely, the Arctic waters, the North Sea, the Celtic Sea, Bay of Biscay and Iberian Coast, and the wider Atlantic. Plastic litter represents approximately 90% of beach litter found in the North Sea, while its share is slightly lower for the Celtic Sea and the Bay of Biscay (78% and 71%, respectively). The Arctic waters have the highest level of plastic litter with approximately 98% of all litter being plastic. The most common litter items found in all regions are plastic fragments, nets, and ropes. There is a general lack of information on the abundance of microplastics in the North East Atlantic (OSPAR, 2017).

The Baltic Sea is the largest expanse of brackish water in the world. Being a relatively shallow semi-enclosed sea, water exchange is limited. Various sources contribute to marine litter, including tourism, wastewater discharge, and river fisheries (Strand et al., 2016). Approximately 70% of the marine litter in the Baltic Sea consists of plastic. The highest density of marine litter along the coasts of the Baltic Sea is documented for the Gulf of Finland, Bothnian Sea, and Northern Baltic Proper, with an average of 280 items per hundred meter. The most common litter items include plastic wraps, bottles, lids as well as cigarette butts (HELCOM, 2018). Moreover, fishing gear and nets are abundant in the Eastern Gotland Basin, Gdansk Basin and Kiel Bay (HELCOM, 2018). Some studies have monitored the concentration of microplastics in the Gulf of Finland, the Belt Sea, and near Stockholm (Setälä et al., 2016; Tamminga et al., 2018; Gewert et al., 2017). The results varied from 0.3-2.1 particles per cubic metre in the Gulf of Finland (Setälä et al., 2016) to 0.04-0.09 particles per cubic metre in the Belt Sea (Tamminga et al., 2018).

Like the Baltic Sea, the Mediterranean Sea is a semi-enclosed sea with limited water exchange. More than 95% of the floating marine litter in the Mediterranean Sea is plastic, while approximately 50% of the seabed litter is plastic (UNEP/MAP and Plan Bleu, 2020). An approximate amount of 730 tonnes of plastic litter is discharged into the Mediterranean every day (UNEP/MAP, 2015). The most dominant marine litter category is single-use plastics with more than 60% of recorded items (Addamo et al., 2017). In terms of microplastics, the registered concentration at the surface of the Mediterranean Sea is approximately 100 thousand items per square kilometre (UNEP/MAP, 2015). However, Van Der Hal et al. (2017) recorded maxima of more than 64 million floating particles per square kilometre.

# 3 Study design and econometric approach

## 3.1 Design of the choice experiment

The goal of this study was to elicit economic values for different marine litter reduction strategies from the perspective of EU citizens. Marine litter reduction strategies can be designed in a variety of ways, depending on the goal of the strategy. These design characteristics can include, for example, the location where the strategy will be implemented, how effective the strategy will be, the environmental benefits of the strategy, and the financial costs involved. A DCE is an ideal tool to determine preferences for different design characteristics, since it allows for a variety of design characteristics to be combined into programmes of measures. The following section describes how the DCE was developed and implemented.

A central element of designing a DCE is to decide on the number of choice alternatives, and to select the attributes and associated levels that describe them. The choice tasks in this study provided respondents with three alternatives. They could choose among two hypothetical alternatives (Program A and Program B) and the opt-out alternative. The hypothetical alternatives described the future state of the marine environment by 2030 with reduced levels of plastic litter in the European seas. The opt-out alternative described the State in 2030 without additional measures to clean up plastics from European seas. Although the MSFD aims to achieve a GES of the marine environment by 2020, we choose a longer period in our DCE design as the GES will not be reached for most of the descriptors, including marine litter (EC, 2018).

The attributes aim at representing dimensions of the future state of the marine environment. These attributes and the associated levels were selected based on a literature review and expert interviews. Pictograms were designed to visually describe each attribute and attribute level. Attributes and pictograms were tested in a series of focus groups at the end of 2019 and the beginning of 2020. Focus group participants included experts, lay public, and students at the Vrije Universiteit Amsterdam, the Netherlands and Kiel University in Germany.

The final set of six attributes used in the DCE is presented in Table A1 in the Appendix to this paper. The first four attributes directly relate to marine plastic pollution, one relates to additional strategies that can reduce plastic pollution and the final attribute is the payment vehicle.

The first attribute is environmental status of the marine environment and directly relates to the MSFD. For marine litter, which consists mostly of plastic litter, GES was defined as in the MSFD's Annex I as "properties and quantities of marine litter [that] do not cause harm to the coastal and marine environment". This includes both animals and plants. The attribute has four possible levels: poor (opt-out), moderate, good, and excellent. The second attribute considers outcome (un)certainty, as there is uncertainty about actually reaching the desired improvement, depending on the extent of the problem and the specific implemented measures. This attribute has three levels: 50%, 75%, and 99% certainty of achieving the environmental status by 2030. Third, an attribute related to the size of the plastic litter was included to reflect that the programmes of measures can target different sizes of plastic litter, depending on the technologies applied. This attribute has four levels, consisting of no removal (opt-out), removing micro plastic, removing macro plastic, and removing both micro and macro plastic. The fourth attribute included was geographic coverage where the measures would be implemented: nowhere (opt-out), in all European seas or solely in the sea(s) bordering the respondent's country. This geographic scope attribute was added to measure to what extent respondents support EU-wide measures or whether they prefer measures in their own marine territories only.

Besides removing plastic litter, additional waste management measures were proposed to prevent plastic from ending up in marine environments and thereby avoid future clean-up of plastic litter. The fifth attribute has five levels: no additional waste management measures (opt-out), improved recycling labeling, improved collection of plastic through a deposit-refund system, a ban of single-use plastic, or all of these three measures applied together.

Since achieving improvements in the environmental status of the European seas comes at a cost, the final attribute is the financial contribution respondents are asked to pay for their preferred alternative. In the EU, waste management, including waste collection, is primarily funded by charging households a waste fee for the costs of waste collection and processing. Information about waste collection and processing costs and fees is very limited. A study assessing waste collection schemes in the capital cities of the 28 Member States in 2015 reports fees ranging from 51 (Vilnius) to 280 (Amsterdam) euros per capita annually, and an average payment across Europe of around 145 euros per capita annually for separate waste collection (BiPRO/CRI, 2015). This information provided the basis for the establishment of the additional waste management fees in the DCE to be paid by a household per month for the next 10 years: no payment (opt-out), €5, €10, €20, €35, and €50. In order to account for the fact that waste collection and management fees are collected annually in some EU Member States, the annual amounts were provided in the DCE next to the monthly payments. For Denmark and Sweden, the Euro amounts were converted to Danish and Swedish Crowns.

Based on these attributes and attribute levels, a D-efficient experimental design (Rose and Bliemer, 2009) was generated with the software NGENE consisting of 24 choice tasks, which were assigned to four blocks, hence resulting in each respondent answering 6 choice tasks. An example choice card is presented in Figure 2.

| | **Program A** | **Program B** | **State in 2030 with no additional measures** |
|---|---|---|---|
| **Environmental status** | 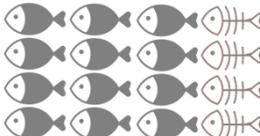 Good | 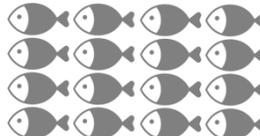 Excellent | 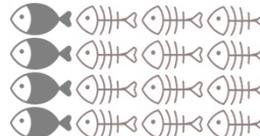 Poor |
| **Certainty** | 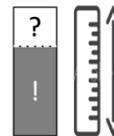 75% certain | 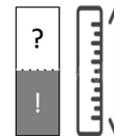 50% certain | 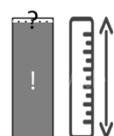 99% certain |
| **Size of plastic litter** | 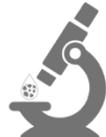 Remove <u>micro</u> plastics | 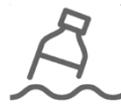 Remove <u>macro</u> plastics | 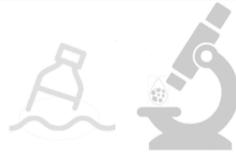 No removal |
| **Geographic coverage** | 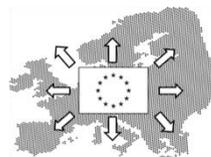 All European Seas | 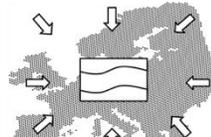 Your Country's Seas | 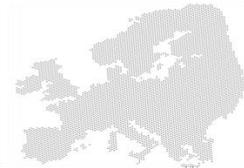 Nowhere |
| **Waste management measure** | 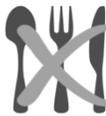 Ban single-use plastics | 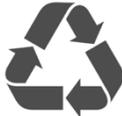 Improved recycling labeling | 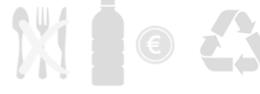 No measures |
| **Additional waste management fee to your household** | 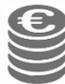 €20 per month (€240 per year) | 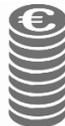 €50 per month (€600 per year) | 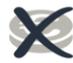 No payment |
| **I prefer:** | ◯ | ◯ | ◯ |

*Figure 2: Example choice card*

The DCE was embedded in an online survey. After asking questions about their socio-demographic background for quota sampling purposes, including age, gender, education and region of residence, respondents were asked if they had ever visited the sea and coast of a European country for leisure. If so, they were asked to list the countries where they had visited the coast or beach. Those that had visited the coast of a European country for leisure in 2019 were then asked more detailed questions regarding the specific sites they visited most frequently, including the number of days they visited, their leisure activities, the importance of specific beach site characteristics for visiting, and beach site quality conditions, such as water clarity, the amount and size of plastic beach litter, and the number of times they saw plastic litter.

This was followed by the DCE section. The choice tasks were introduced using a short text describing that the EU aims at preventing and cleaning up marine litter in European seas to achieve GES by 2030. Respondents were informed that 60-80% of all marine litter consists of plastic litter that is left behind by people on the beach or ends up in the sea through rivers, storm water runoff, waves, and wind. Also, the difference between macro litter (larger items such as plastic bottles, parts of fishing nets, toys and plastic bags) and micro litter (small fragments of larger plastic items or personal care products like cleansers and scrubbing powders that are invisible to the naked eye) was explained. Respondents were shown Table A1 to familiarize them with the attributes and their different levels, as well as the example choice card shown in Figure 2 accompanied by explanatory text. After the DCE, they were asked a few debriefing questions about the DCE, and a number of additional questions about their socio-economic background, in particular their household income.

The DCE and the other questions in the survey were developed, screened and tested over a period of approximately 12 months. The draft DCE design was also presented for comments and feedback to the large group of marine litter experts collaborating with the authors of this paper in the Horizon2020 project CLAIM during the annual project meeting in Copenhagen in October, 2019. The survey was translated into the official languages of the eight countries where the survey was implemented by the Norwegian survey company Norstat in the months July and August, 2020. Norstat manages and has access to representative household panels in the mentioned European countries from which the respondents in this study were sampled. The design of the survey was also discussed extensively with the marketing experts working for Norstat before it was launched online. Before final implementation, the online survey was pre-tested using 100 respondents in two of the eight countries (i.e. Italy and the Netherlands), on the basis of which conditional logit and mixed logit choice models were estimated to assess their overall goodness of fit to the data and the statistical significance and direction of influence of the individual choice attributes on choice behavior. Special attention was paid to the presence of possible protest behavior. All attributes behaved as expected in the estimated choice models and no signs of protest behavior were detected. In view of these results, no further modifications were made to the design and these responses were retained in the final sample.

### 3.2 Econometric model

The econometric modelling of the choice data is based on well-established random utility maximization (RUM) models (McFadden, 1973). The standard random utility model is specified in equation (1), where an individual $i$ derives utility $U_{ij}$ from alternative $j$ if alternative $j$ is chosen. In this model, the utility $U_{ij}$ is a function of a deterministic (observable) and a stochastic (unobservable) part. $V_{ij}$ is the observable vector of the $k$ attributes applied in the DCE here and measured by their levels as explanatory variables. The $V_{ij}$ vector can more generally be assumed to be a linear additive function of $Z$, a vector of the $k$ observable attributes varying across individuals $i$ and alternatives $j$, and multiplied by $\beta$, a vector of $k$ coefficients and their possible interaction with individual respondent characteristics. The unobservable stochastic component is captured in the utility function (1) by the random error term $\varepsilon$ which is IID extreme value type I distributed within each choice (Train, 2003).

$$U_{ij} = V_{ij} + \varepsilon_{ij} = \beta' Z'_{ij} + \varepsilon_{ij} \qquad (1)$$

We employ a mixed logit (MXL) model, which allows us to account for heterogeneous preferences, specified in the random parameters associated with the choice attributes. The adjusted utility function of the MXL model is presented in equation (2), where the first part is identical to equation (1), but the vector of coefficients $\beta_i$ is now randomized over the sample population, as shown in equation (3). Here, the vector of coefficients $\beta_i$ is decomposed into $\mu_i$, the sample population mean utility attached to a specific choice attribute, and $\theta$ the standard deviation, which captures the individual preferences relative to the average population preference, and is typically a mixing distribution function for $\beta_i$. In this study, the distributions of the random parameters related to the choice attributes are all assumed to have a normal distribution, except the price attribute, which is assumed to have a lognormal distribution, as this generated the best model fit.

$$U_{ij} = V_{ij} + \varepsilon_{ij} = \beta'_i Z'_{ij} + \varepsilon_{ij} \quad (2)$$

$$\beta'_i Z'_{ij} = \mu_i Z'_{ij} + \theta Z'_{ij} \quad (3)$$

## 4 Results

### 4.1 Descriptive statistics

In each of the eight sampled countries we gathered data for approximately 1,000 individuals aged 18 or above. Table 1 provides the descriptive statistics for some of the main socio-demographic characteristics of the samples and the populations from which they were drawn. The sample in each country represent the underlying target populations well. Only small differences exist between the sample and the target population, most importantly for Greece. Slightly more women than men responded to the survey in most of the countries, which corresponds closely to the target population distribution. More men than women participated in Greece. The samples' average age lies in the late 40's, except in Germany and Greece, and is overall relatively lower than that of the target populations (average age of early 50's). This small difference in average age can be explained by the fact that the survey was administered online which was therefore expected to reach a slightly younger population. The distribution of net monthly household income across different income groups matches the distributional patterns of the target population fairly well, but is skewed to the higher income categories in all sampled countries, with the highest relative difference for the Greek and Swedish sample. A higher income in sampled populations is not uncommon in online surveys (e.g., Lindhjem and Navrud, 2011). The higher income levels in the sampled populations are reflected in the higher share of respondents that have completed an educational level of at least upper secondary education (ISCED 3), with again differences being most prominent for Greece and Sweden.

As shown in Table 2, the survey also included several questions asking respondents for their sea or beach visitation behaviour in their own or another European country. The majority of respondents had visited their own as well as another European country's sea and coast. Most French and Greek respondents, over one in five respondents, indicated they only visit their own beaches and seas. Note that the percentages of respondents visiting beaches and coastal zones in Table 2 do not add up to 100% because they exclude the share of respondents who indicated they never visited a European sea or coast. For some country samples this share is considerable, and is highest for Italy (27%) and lowest for Sweden (4%). If respondents had visited a European coastal site in 2019, which was the case for more than 50% of the whole survey sample, several follow-up questions were asked addressing their perception of the site they visited most often. The absence of beach litter is perceived most important for respondents living in France, Italy and Greece. Greek, Dutch and Swedish respondents reported to see litter on average most often, while the most amount of litter was reported on average by the same respondents in France and Italy. The largest pieces of plastic litter were observed in Italy and Greece in the Mediterranean Sea, and Germany and the Netherlands who border the North Sea. In the case of Germany this also relates to the Baltic Sea. Applying the Kruskal-Wallis test shows that the observed differences between countries, albeit small in some cases, are statistically significant in terms of frequency, amount and size of perceived litter, as well as the importance attached to not having any litter on beaches. Test results are available from the authors upon request.

Table 1: Descriptive statistics of the samples' socio-demographic background characteristics

| | Denmark | | Sweden | | Estonia | | Germany | | The Netherlands | | France | | Italy | | Greece | |
|---|---|---|---|---|---|---|---|---|---|---|---|---|---|---|---|---|
| | Sample | Whole population | Sample | Whole population | Sample | Whole population | Sample | Whole population | Sample | Whole population | Sample | Whole population | Sample | Whole population | Sample | Whole population |
| **Share male (%)** | 48.90 | 49.76 | 49.40 | 50.72 | 45.51 | 47.22 | 48.6 | 49.35 | 49.10 | 49.65 | 47.08 | 48.33 | 48.0 | 48.68 | 52.30 | 48.56 |
| **Average age (years)** | 48.42 (17.37) | 51.50 | 49.22 (17.98) | 51.46 | 47.58 (16.72) | 51.55 | 40.18 (16.41) | 52.72 | 47.21 (17.09) | 51.71 | 48.04 (16.42) | 52.70 | 48.92 (15.56) | 53.60 | 42.26 (13.95) | 53.13 |
| **Average monthly household income (€)** | 3,610.33 (2,289.24) | 2,813.25 | 3,502.10 (2,110.54) | 2,311.08 | 1,563.35 (1,094.46) | 979.25 | 2,659.06 (1,610.32) | 2,155.92 | 2,822.40 (1,834.21) | 2,239.33 | 2,360.89 (1,448.94) | 2,120.50 | 1,965.49 (1,262.05) | 1,600.67 | 1,473.35 (1,423.60) | 754.08 |
| **Average household size (persons)** | 2.24 (1.31) | 2.00 | 2.24 (1.19) | 2.00 | 2.25 (1.29) | 2.20 | 2.27 (1.49) | 2.00 | 2.72 (3.13) | 2.00 | 2.50 (2.46) | 2.20 | 2.97 (3.06) | 2.30 | 3.14 (2.75) | 2.60 |
| **Share higher education (ISCED 6-8) (%)** | 31.90 | 25.30 | 28.10 | 38.60 | 35.10 | 31.72 | 21.90 | 25.06 | 39.94 | 26.61 | 33.70 | 24.18 | 15.20 | 12.27 | 61.80 | 27.80 |
| **N** | 1,000 | | 1,000 | | 1,002 | | 1,000 | | 1,004 | | 1,009 | | 1,000 | | 1,000 | |

Population data are from Eurostat and refer to 2018 (household size and income) and 2019 (education, gender, and age) (https://ec.europa.eu/eurostat/web/main/data/database; last retrieved September 2020). Sample standard deviations are presented between brackets.

**Table 2: Recreational visitation behaviour to European seas and public perception of marine litter**

|  | Denmark | Sweden | Estonia | Germany | The Netherlands | France | Italy | Greece |
|---|---|---|---|---|---|---|---|---|
| Sample size | N=1,000 | N=1,000 | N=1,002 | N=1,000 | N=1,004 | N=1,009 | N=1,000 | N=1,000 |
| Share visiting European sea(s) in 2019 for recreation (%) | 73.78 | 61.92 | 75.32 | 56.14 | 67.81 | 67.03 | 67.81 | 73.85 |
| **Share visiting (%)** | | | | | | | | |
| Own country's sea only | 0.00 | 3.00 | 11.89 | 8.02 | 8.57 | 22.86 | 12.20 | 21.92 |
| Other EU countries' seas only | 7.51 | 13.71 | 14.19 | 20.76 | 22.63 | 11.03 | 8.20 | 13.11 |
| Own and other EU countries' seas | 80.58 | 78.88 | 58.84 | 58.27 | 56.03 | 46.92 | 52.30 | 45.65 |
| **For the most often visited coastal sites in 2019[a]** | | | | | | | | |
| Importance of absence of plastic beach litter (1-4)[b] | 2.98 (3) | 3.22 (3) | 3.35 (4) | 3.32 (3) | 2.91 (3) | 3.44 (4) | 3.45 (4) | 3.63 (4) |
| Frequency of seeing beach litter (1-4)[c] | 1.85 (2) | 2.10 (2) | 1.65 (2) | 1.90 (2) | 2.07 (2) | 1.99 (2) | 1.96 (2) | 2.05 (2) |
| Amount of beach litter (1-4)[d] | 2.04 (2) | 2.09 (2) | 2.00 (2) | 2.15 (2) | 2.15 (2) | 2.22 (2) | 2.25 (2) | 2.16 (2) |
| Average size of beach litter (1-4)[e] | 2.27 (2) | 2.21 (2) | 2.15 (2) | 2.28 (2) | 2.26 (2) | 2.10 (2) | 2.22 (2) | 2.19 (2) |

[a] For all questions mean values are presented with median values between brackets. The last three questions included the option 'don't know or don't remember' which was disregarded when calculating the mean and median values.
[b] How important was the absence of plastic litter on the beach in choosing how often you went there in 2019? (1=not important at all; 4=very important).
[c] How often did you see plastic litter on the beach or in the water? (1=never; 4=every time).
[d] How would you describe the amount of plastic litter on the beach or in the water? (1=no pieces; 4=Very many pieces).
[e] How would you describe the size of the most noticeable plastics on the beach or in the water? (1=very small; 4=very large).

## 4.2 Preferences for marine plastic litter reduction

Figure 3 shows the choice behavior for the three alternatives, including the opt-out, across the eight samples. Alternatives A and B were almost equally chosen within each sample and across the samples, with differences within countries being not much higher than a few percentage points. A and B were chosen equally often in the Netherlands and France. Alternative A was chosen between 44% (Netherlands and France) and 50% (Denmark) of all the choices, while Alternative B was chosen between 43% (Estonia) and 46% (Denmark and Sweden) of all the choices. More pronounced differences were found between samples for the opt-out responses. Opt-out responses are substantially lower in Denmark (4%) and Sweden (5%) compared to France (12%), Estonia, and the Netherlands (11%).

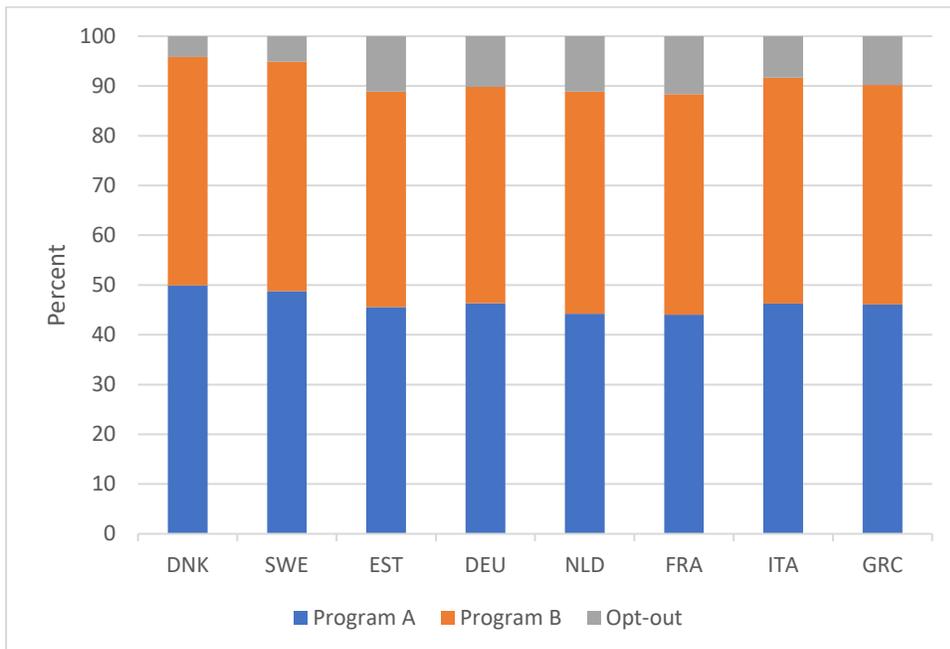

*Figure 3: Choice behaviour across the samples*

Respondents who consistently chose the opt-out were examined in more detail. A total of 353 respondents were identified as potential protestors and dropped from further analysis, ranging between 2.9% (The Netherlands) and 6.0% (France) of a country's sample. Protestors were identified as those respondents who always chose the opt-out and indicated in a follow-up question that either the government is responsible for the clean-up of marine litter or the polluter should pay.

The remaining choices were then regressed on the choice attributes using Stata 16. The results of the MXL choice models are presented in Table 3. For all countries, most of the estimated standard deviations of the random coefficients are significant, implying significant unobserved taste heterogeneity between respondents, and justifying the use of the MXL model. The results for the attributes-only models are presented here because not many covariates appeared to significantly influence choice behavior or the utility attached to the various choice attributes (further discussed below).

The results for the alternative specific constant (ASC) related to the opt-out has the expected negative sign and is statistically highly significant across the samples in all the countries, indicating that respondents prefer a change instead of no change irrespective of the levels of the different choice attributes, and the opt-out alternative represented the option with the least utility. Also, the parameter estimates for the cost attribute, the waste management fee, are consistently negative and highly significant (at the 1% level) across all eight country samples, indicating that the presented trade-offs between improvements in the marine environment (benefits) and the price (costs) respondents are asked to pay for this worked. Price sensitivity is highest in Greece and the Netherlands and lowest in Estonia.

Compared to the reference level (moderate), the average respondent's utility increases significantly with a good environmental status and even more so with an excellent one. The differences between the estimated coefficient estimates for good and excellent status are statistically significant based on the Wald test. Similar results are found for the certainty of reaching the proposed improvement in environmental status, albeit less pronounced when examining the differences between the coefficient estimates for a 75% and 99% outcome certainty. Only the difference in Estonia between 75% and 99% is not statistically significant based on the Wald test. Moreover, a certainty level increase from 50% to 75% is not statistically significant in the Netherlands and France. In these latter two countries, respondents only value the environmental status improvement, and are willing to pay for it, when this improvement is 99% certain. Further, in all samples the public significantly prefers to address marine plastic removal in all European seas instead of in the respondent's country of origin only, and is willing to pay extra for this.

When it comes to which plastics should be removed from the sea, a remarkable finding is that removing both plastic sizes, micro and macro, significantly increases utility compared to removing macro plastic only in all countries, except for Greece. Greek respondents prefer to remove macro plastics only, not micro plastics. Examining the different ways of avoiding litter under the additional waste management measures indicates a strong preference for all the measures that were proposed in the DCE. The coefficient estimates for individual measures are all negative, compared to the baseline of implementing all measures, but not always statistically significant.

Turning to the random effects, significant preference heterogeneity is detected most importantly related to the highest level of improvement of the environmental status across all European seas and the price respondents are asked to pay for it. The random terms for micro plastics and the labelling of recyclable products are also both consistently significant across all eight country samples, indicating that preferences among individual respondents vary significantly over these two attributes.

Finally, trying to capture some of that unobserved preference heterogeneity, we also estimated extended choice models for each country sample by including possible sources of the preference heterogeneity. Covariates, systematically accounted for across the different countries, included interaction terms between (i) whether or not respondents had visited coastal areas or beaches in 2019 and the environmental status of a sea in 2030; (ii) respondent gender and the certainty of reaching an environmental status; (iii) respondent age and the geographical coverage of the proposed policy alternative; and (iv) respondent income and the cost attribute. The results of the MXL choice models including covariates are included in the appendix to this paper (Table A2). Although the covariates did not significantly improve the model fit, the interaction effects reveal interesting patterns. Most importantly, the interaction between the waste management fee and household income generates, as theoretically expected, a (small) significant positive effect in five of the eight EU countries (Denmark, Sweden, Germany, the Netherlands and Greece), indicating that public WTP depends on ability to pay. Hence, the higher a respondent's household income, the higher on average their WTP. The coefficient estimate in Italy is positive but just not significant at the 10% level, our cut-off point for identifying statistically significant effects. Similarly, as theoretically expected, whether or not someone is considered a 'user[2]' of marine and coastal amenities also significantly affects the utility respondents attach to improving the environmental status of marine ecosystems in a positive way in four countries (Denmark, Germany, France and Greece). In most countries being a 'user' significantly adds value to both an improvement from moderate to good and from moderate to excellent status, with a higher coefficient for excellent than good, as expected, except in Germany where a 'good' status is considered good enough and improving the environmental status further to excellent does not yield any extra value. Gender effects could be detected in Estonia, Italy and France, although results are only marginally significant. Compared to an outcome certainty level of 50%, female respondents in Estonia and Italy preferred a level of 75% compared to male respondents, while female respondents in France favoured an even higher outcome certainty level of 99%. A respondent's age appeared to have a significant impact on the utility respondents attached to the geographical coverage of the policy alternatives. However, the impact of age in the extended choice models are mixed between countries, with older respondents in Sweden, Estonia and Greece preferring the quality of their own country's sea to be improved, while older respondents in the Dutch sample had a significantly higher preference for a European instead of local coverage of the quality improvements.

---

[2] A respondent who visited coastal areas or beaches in the previous year.

**Table 3: Estimated mixed logit choice models across the eight European countries**

|  | Denmark | Sweden | Estonia | Germany | Netherlands | France | Italy | Greece |
|---|---|---|---|---|---|---|---|---|
|  | Coef. | Coef. | Coef. | Coef. | Coef. | Coef. | Coef. | Coef. |
|  | (Std.err.) | (Std.err.) | (Std.err.) | (Std.err.) | (Std.err.) | (Std.err.) | (Std.err.) | (Std.err.) |
| **ASC$_{OPT}$** | -3.8945*** | -3.4428*** | -3.2588*** | -2.9132*** | -2.6714*** | -2.8842*** | -3.0385*** | -3.4590*** |
|  | (0.1641) | (0.1451) | (0.1309) | (0.1257) | (0.1040) | (0.1149) | (0.1195) | (0.1209) |
| **ENVIRONMENTAL STATUS (BASELINE CATEGORY: MODERATE)** | | | | | | | | |
| **GOOD** | 0.8179*** | 0.8139*** | 0.4094*** | 0.6023*** | 0.3339*** | 0.3161*** | 0.3704*** | 0.2555*** |
|  | (0.0681) | (0.0709) | (0.0653) | (0.0651) | (0.0545) | (0.0582) | (0.0594) | (0.0532) |
| **EXCELLENT** | 1.1871*** | 1.2779*** | 0.7771*** | 1.1874*** | 0.6997*** | 0.7422*** | 0.7802*** | 0.5552*** |
|  | (0.0809) | (0.0894) | (0.0776) | (0.0815) | (0.0658) | (0.0714) | (0.0718) | (0.0601) |
| **CERTAINTY LEVEL (BASELINE CATEGORY: 50%)** | | | | | | | | |
| **75%** | 0.2267*** | 0.2689*** | 0.1998*** | 0.1585** | 0.0129 | 0.0706 | 0.1365** | 0.1178** |
|  | (0.0598) | (0.0632) | (0.0635) | (0.0615) | (0.0534) | (0.0566) | (0.0579) | (0.0523) |
| **99%** | 0.3783*** | 0.4947*** | 0.2623*** | 0.3247*** | 0.2495*** | 0.2752*** | 0.3389*** | 0.2611*** |
|  | (0.0624) | (0.0655) | (0.0633) | (0.0625) | (0.0565) | (0.0586) | (0.0605) | (0.0529) |
| **PLASTIC SIZE (BASELINE CATEGORY: MACROPLASTIC)** | | | | | | | | |
| **MICROPLASTIC** | 0.0743 | -0.0839 | -0.0021 | -0.0108 | -0.0614 | -0.0499 | 0.0791 | -0.1601*** |
|  | (0.0630) | (0.0659) | (0.0655) | (0.0645) | (0.0559) | (0.0613) | (0.0619) | (0.0564) |
| **BOTH PLASTIC SIZES** | 0.4466*** | 0.4854*** | 0.2233*** | 0.3519*** | 0.2082*** | 0.2118*** | 0.4450*** | 0.0289 |
|  | (0.0648) | (0.0679) | (0.0684) | (0.0677) | (0.0551) | (0.0589) | (0.0650) | (0.0530) |
| **GEOGRAPHIC LOCATION (BASELINE CATEGORY: OWN COUNTRY)** | | | | | | | | |
| **EU** | 0.7122*** | 0.7166*** | 0.4672*** | 0.5864*** | 0.3881*** | 0.2859*** | 0.3144*** | 0.2357*** |
|  | (0.0586) | (0.0603) | (0.0584) | (0.0585) | (0.0498) | (0.0511) | (0.0517) | (0.0439) |
| **WASTE MANAGEMENT MEASURE (BASELINE CATEGORY: ALL)** | | | | | | | | |
| **LABEL** | -0.1829** | -0.3522*** | -0.1377* | -0.2465*** | -0.0576 | -0.0955 | -0.1077 | -0.1750*** |
|  | (0.0726) | (0.0781) | (0.0807) | (0.0792) | (0.0664) | (0.0727) | (0.0746) | (0.0658) |
| **REFUND** | -0.1747** | -0.2516*** | -0.0999 | -0.0796 | -0.0361 | -0.0215 | -0.0339 | -0.0309 |
|  | (0.0778) | (0.0793) | (0.0786) | (0.0773) | (0.0684) | (0.0732) | (0.0737) | (0.0671) |
| **BAN** | -0.2895*** | -0.3357*** | -0.2627*** | -0.1728** | -0.2262*** | -0.2094*** | -0.2291*** | -0.1147* |
|  | (0.0794) | (0.0862) | (0.0835) | (0.0806) | (0.0693) | (0.0741) | (0.0767) | (0.0677) |
| **COST** | -3.2586*** | -3.3163*** | -2.8156*** | -3.0668*** | -3.4839*** | -3.2926*** | -3.4147*** | -3.5290*** |
|  | (0.0836) | (0.0892) | (0.0784) | (0.0829) | (0.093) | (0.0954) | (0.0925) | (0.0959) |

| (CONT.) | Denmark | Sweden | Estonia | Germany | Netherlands | France | Italy | Greece |
|---|---|---|---|---|---|---|---|---|
| | Coef. | Coef. | Coef. | Coef. | Coef. | Coef. | Coef. | Coef. |
| | (Std.err.) | (Std.err.) | (Std.err.) | (Std.err.) | (Std.err.) | (Std.err.) | (Std.err.) | (Std.err.) |
| **SD** | | | | | | | | |
| **ENVIRONMENTAL STATUS (BASELINE CATEGORY: MODERATE)** | | | | | | | | |
| **EXCELLENT** | 0.3997** | 0.8696*** | 0.6846*** | -0.6122*** | -0.7878*** | -0.7544*** | -0.7282*** | -0.5960*** |
| | (0.1553) | (0.1152) | (0.1345) | (0.1286) | (0.1011) | (0.1134) | (0.1140) | (0.1049) |
| **GOOD** | 0.0068 | 0.0140 | -0.1149 | 0.0511 | 0.0071 | -0.0063 | -0.0367 | -0.0159 |
| | (0.1338) | (0.1821) | (0.3609) | (0.3026) | (0.1709) | (0.1788) | (0.1623) | (0.2859) |
| **CERTAINTY LEVEL (BASELINE CATEGORY: 50%)** | | | | | | | | |
| **75%** | 0.0136 | 0.0131 | 0.0011 | -0.0083 | -0.0024 | -0.0004 | 0.0038 | -0.0026 |
| | (0.1218) | (0.1051) | (0.1188) | (0.1263) | (0.1087) | (0.1182) | (0.1403) | (0.0943) |
| **99%** | 0.0545 | 0.0789 | 0.0027 | -0.0174 | -0.3925*** | -0.2070 | 0.3540** | -0.0469 |
| | (0.1713) | (0.2251) | (0.2140) | (0.2553) | (0.1324) | (0.2065) | (0.1757) | (0.1358) |
| **PLASTIC SIZE (BASELINE CATEGORY: MACROPLASTICS)** | | | | | | | | |
| **MICROPLASTICS** | 0.6386*** | 0.6916*** | 0.6484*** | 0.5996*** | 0.4419*** | -0.6537*** | 0.7140*** | 0.5763*** |
| | (0.1239) | (0.1247) | (0.1316) | (0.1311) | (0.1321) | (0.1158) | (0.1109) | (0.1076) |
| **BOTH PLASTIC SIZES** | 0.3632* | -0.1609 | 0.6067*** | -0.6027*** | 0.0795 | -0.1503 | -0.5126*** | -0.0009 |
| | (0.2024) | (0.4849) | (0.1507) | (0.1433) | (0.3808) | (0.3749) | (0.1439) | (0.1663) |
| **GEOGRAPHIC LOCATION (BASELINE CATEGORY: OWN COUNTRY)** | | | | | | | | |
| **EU** | 0.8476*** | 0.8459*** | 0.7519*** | 0.8085*** | -0.7038*** | -0.7089*** | 0.7279*** | 0.5314*** |
| | (0.0861) | (0.0912) | (0.0934) | (0.0873) | (0.0758) | (0.0849) | (0.0832) | (0.0803) |
| **WASTE MANAGEMENT MEASURE (BASELINE CATEGORY: ALL)** | | | | | | | | |
| **LABEL** | -0.6301*** | 0.8151*** | 0.9499*** | -0.8794*** | -0.4511** | -0.7728*** | -0.8790*** | -0.5699*** |
| | (0.1713) | (0.1539) | (0.1597) | (0.1496) | (0.1953) | (0.1526) | (0.1429) | (0.1534) |
| **REFUND** | -0.0065 | 0.0152 | 0.1619 | -0.0104 | 0.0041 | 0.0481 | -0.0036 | -0.0057 |
| | (0.1625) | (0.1950) | (0.2038) | (0.1798) | (0.1555) | (0.1818) | (0.1549) | (0.1496) |
| **BAN** | 0.1757 | 0.8711*** | 0.5527*** | 0.2403 | 0.3068 | -0.3582* | 0.5649*** | -0.2882 |
| | (0.2769) | (0.1443) | (0.1793) | (0.2908) | (0.2442) | (0.2132) | (0.1656) | (0.2240) |
| **COST** | 1.3801*** | 1.3919*** | 1.3440*** | 1.4937*** | 1.6666*** | 1.6472*** | 1.5354*** | 1.8108*** |
| | (0.0655) | (0.0785) | (0.0584) | (0.0669) | (0.0773) | (0.0744) | (0.0683) | (0.0792) |

| (CONT.) | Denmark | Sweden | Estonia | Germany | Netherlands | France | Italy | Greece |
|---|---|---|---|---|---|---|---|---|
| | Coef. (Std.err.) | Coef. (Std.err.) | Coef. (Std.err.) | Coef. (Std.err.) | Coef. (Std.err.) | Coef. (Std.err.) | Coef. (Std.err.) | Coef. (Std.err.) |
| **NUMBER OF OBS** | 17,208 | 17,388 | 17,100 | 16,902 | 17,550 | 16,992 | 17,424 | 17,352 |
| **NUMBER OF RESPONDENTS** | 956 | 966 | 950 | 939 | 975 | 944 | 968 | 964 |
| **LR CHI2(10)** | 1068.00 | 1037.48 | 1459.25 | 1472.76 | 1404.93 | 1499.74 | 1244.89 | 1651.94 |
| **PROB > CHI2** | 0.0000 | 0.0000 | 0.0000 | 0.0000 | 0.0000 | 0.0000 | 0.0000 | 0.0000 |
| **LOG LIKELIHOOD** | -3445.826 | -3692.827 | -3991.496 | -3875.640 | -4447.509 | -4233.746 | -4218.959 | -4186.429 |
| **AIC** | 6937.653 | 7431.654 | 8028.991 | 7797.281 | 8941.017 | 8513.493 | 8483.918 | 8418.859 |
| **BIC** | 7115.975 | 7610.215 | 8207.168 | 7975.19 | 9119.792 | 8691.524 | 8662.527 | 8597.73 |

*Note: *= statistically significant at the 10% level; **= 5% level; and *** = 1% level.*

## 4.3 Willingness to pay for marine litter reduction

Based on the results of the MXL model presented in Table 3, marginal WTP values were calculated for each attribute level (see Table A3 in the Appendix). These marginal WTP values reflect what an average household in the sample would be willing to pay annually in each of the eight countries. The baseline for these values is the same reference group that was used in the MXL model.

In Figure 3, the marginal annual WTP values and their 95% confidence intervals for four attributes and five levels are shown for each country in our sample. The five levels illustrated in Figure 3, namely good environmental status, excellent environmental status, applying the measures at EU instead of individual country scale, removing both macro and micro plastics, and having an improved environmental state with a 99% certainty level, are consistently significant in all countries except for Greece where removing both plastic sizes was not significantly different from removing only macro plastic.

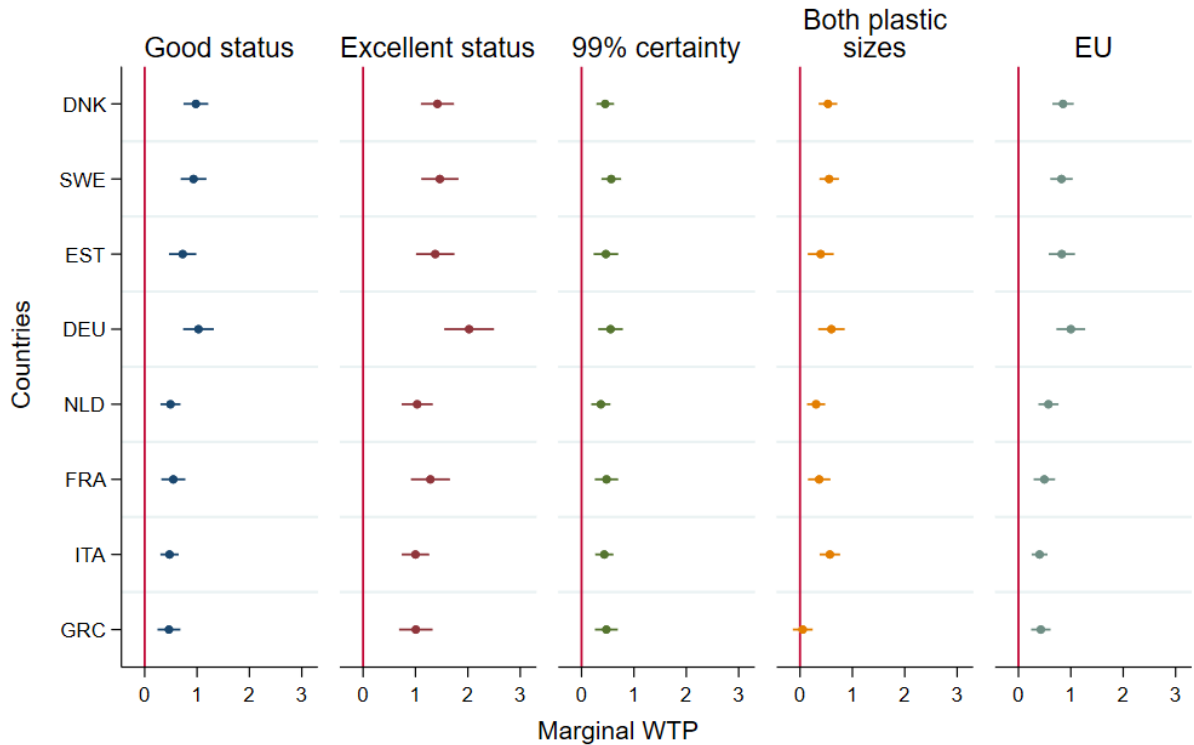

*Figure 3: Marginal WTP estimates (in € per year per household)*

Going from moderate to excellent environmental status has the highest marginal WTP value in all countries. Notably, Germany, Sweden and Denmark exhibit the highest marginal WTP values for all the attribute levels, while Italy, France and Greece present the lowest values.

As shown in Table 4, the marginal WTP values can be combined to estimate the (non-marginal) compensating surplus of welfare measures for achieving GES as specified in the MSFD in terms of marine litter reduction. We specify two policy scenarios and compare them with a situation where the environmental quality in 2030 would be moderate, measures would be implemented in individual countries instead of across all European seas, and the focus would be on the removal of macro plastic litter only. The two policy scenarios capture the removal of micro and macro plastic litter and aim to reach GES with 99% certainty. The two scenarios only differ in terms of their geographical coverage. While scenario 1 captures the welfare effects when measures are taken to reach GES in the sea bordering an individual country, scenario 2 covers measures taken in all European seas. As all countries consistently favored alternatives that cover all European seas, the compensation surplus is therefore higher in scenario 2. The differences in welfare gains between scenarios 1 and 2 range between 27% (Italy) and 45% (Germany).

**Table 4: Welfare measures for MSFD policy measures**

|  | Scenario 1<br>*Implementing the MSFD for marine litter in individual EU countries* | Scenario 2<br>*Implementing the MSFD for marine litter at EU scale* |
|---|---|---|
| Environmental status[a] | Good | Good |
| Certainty | 99% | 99% |
| Size of plastic litter | Micro and macro | Micro and macro |
| Geographic coverage | Own country | EU |
|  | Per household (in € per year)[b] | Per household (in € per year)[b] |
| Denmark | 2.0<br>(1.4 - 2.5) | 2.8<br>(2.0 - 3.6) |
| Sweden | 2.1<br>(1.4 - 2.7) | 2.9<br>(2 .0- 3.7) |
| Estonia | 1.6<br>(0.8 - 2.3) | 2.4<br>(1.4 - 3.4) |
| Germany | 2.2<br>(1.4 – 3.0) | 3.2<br>(2.1 - 4.2) |
| The Netherlands | 1.2<br>(0.6 - 1.7) | 1.7<br>(1.0 - 2.5) |
| France | 1.4<br>(0.7 - 2.1) | 1.9<br>(1.0 - 2.8) |
| Italy | 1.5<br>(0.9 – 2.0) | 1.9<br>(1.2 - 2.6) |
| Greece | 0.9<br>(0.5 - 1.4) | 1.4<br>(0.7 – 2.0) |

[a] Compared to a moderate status in 2030 where only macro litter is removed in individual countries.
[b] In brackets are the 95% confidence intervals.

# 5  Discussion and conclusion

Plastic pollution is currently considered one of the most widespread problems affecting the marine environment requiring policy action. The MSFD, the European Union's main legal instrument to protect the marine environment, requires Member States to reduce the amount of marine litter so that it does not cause harm to the coastal and marine environment. This paper adds to the limited empirical evidence base about the non-market benefits of such measures to society. We use a DCE and focus explicitly on plastic litter, distinguishing between micro and macro plastic litter, covering four European seas and eight countries. The trade-off between the benefits of reaching GES and the associated costs are conceptualized in the DCE in terms of six attributes: environmental status, outcome certainty, the size of plastic litter, geographic coverage of the necessary measures, the types of additional waste management measures, and costs in terms of additional waste management fees. Overall, the samples represent the larger country populations from which they were drawn well in terms of socio-demographic background characteristics. Due to thorough pre-testing, the number of protest responses is relatively low in this study and provides credibility to the results reported here.

The results show that European citizens have strong preferences to improve the environmental status of the marine environment by removing both micro and macro plastic litter. Around 90 percent of the respondents agreed to pay for extra measures to improve the environmental status of European seas, either their own or all European seas. When asking the public, they favor a pan-European approach and citizens in the eight countries are also willing to pay extra to improve the environmental status across all European seas compared to just the sea(s) bordering the country in which they live. Similarly, in all the sampled countries, respondents exhibited a preference for implementing a mix of measures instead of picking one only, like a ban on single use plastics, a deposit-refund or better labelling for recycled plastics. As expected, public WTP depends on their ability to pay and we, therefore, detect significant price sensitivity across European countries.

To put these results a bit more in perspective, we compare the welfare effects of achieving GES in terms of marine litter reduction to solid waste collection and processing fees paid by EU households. In the 28 capital cities of the European Union these fees ranged from 51 (Vilnius) to 280 (Amsterdam) euros per capita annually in 2015 (BiPRO/CRI, 2015). Using an average annual payment of 145 euros per capita and an average number of persons per household of two (see Table 1 above), respondents in this study are willing to pay about 1 percent more on top of their current waste collection and processing fees to achieve GES and manage macro and micro litter, compared to a moderate environmental status where only macro litter is reduced. This provides policy makers with important value cues for the social benefits of implementing the MSFD to reach GES in terms of plastic litter, when asking those most likely picking up the bill for implementing the MSFD.

The results suggest similarities and differences in the estimated welfare measures for the different countries. For countries bordering the Baltic Sea, the results for Denmark, Sweden and Germany are similar, while the mean WTP values for achieving GES in the own country's sea are much lower in Estonia. However, the 95% confidence intervals calculated for each of the welfare measures overlap, indicating that the WTP values are not statistically significantly different from each other. For countries situated along the North Sea, the mean WTP values are higher for Germany than for the Netherlands, but this difference is not significant. For countries bordering the Mediterranean Sea, mean WTP for achieving GES in the own country's sea in Italy is 50% higher than in Greece. However, differences between France, Italy and Greece are not significant, which also makes these results informative for decision-makers in other European countries bordering the Mediterranean Sea not included in our samples. Future research could formally test the transferability of the results to countries not covered in this analysis.

Comparing our results to those reported in other studies is not straightforward. Although many stated preference studies exist that value marine ecosystem services, only a handful DCE's focus explicitly on marine litter or plastics. Outside Europe, in the study by Davis et al. (2019), beach cleanliness in Australia was most highly valued by respondents at 51.1 AUD (approx. 32 euros) per unit decrease in litter per square meter, which is difficult to compare to our results given the spatial units used. Furthermore, a study by Schuhmann et al. (2016) revealed that visitors to beaches of Barbados are willing to pay an additional 308 USD (approx. 254 euros) per night to stay at a hotel where the beach has no litter compared to a beach with 5 litter pieces per 25 meter. Other studies estimated WTP for beach cleanliness and improved water quality in terms of WTP per beach visit (Birdir et al. 2013; Blakemore and Williams 2008; Blakemore et al. 2002; Leggett et al. 2018; Loomis and Santiago 2013; Shen et al. 2019; Talpur et al. 2018). WTP estimates in these studies range between 0 and 10 USD (approx. 8 euros) per person and trip (Stöver et al. 2020). An exception is Loomis and Santiago (2013) who valued the removal of beach litter in Puerto Rico at 98 USD (approx. 81 euros) per visitor day. The study by Choi and Lee (2018) is the only one focussing explicitly on micro plastic and is closest to ours, also in terms of welfare estimates. They estimated an annual mean WTP of 2.83 USD (approx. 2.34 euros) per household to remove micro plastics in the ocean of the metropolitan area of Seoul. In Europe, Brouwer et al. (2017) were among the first to distinguish in their DCE design between different types and sources of marine litter, and found a positive

WTP for the removal of plastic litter washed ashore on beaches along three different European seas (North Sea, Black Sea, Mediterranean Sea) varying between 0.65 and 8.25 euro per beach visitor per year, and a slightly lower WTP between 0.40 and 7.05 euro per beach visitor per year for the removal of cigarette butts left behind by beach visitors self. Aanesen et al. (2018) valued the visual intrusion of fish farms and tourism facilities along the coast of Arctic Norway, and reported an average WTP value varying between 123-167 USD (approx. 101-238 euros) per household per year to avoid a 50% increase in litter on the beach. This is substantially higher than the values reported in our study. One of the few studies to investigate the WTP of removing both micro and macro plastic is Abate et al. (2020) focusing on marine plastic pollution in the archipelago of Svalbard. They estimated an average annual WTP value to reduce plastic litter around the archipelago of Svalbard of 642 USD (approx. 530 euros) per household which is also much higher compared to our estimates.

The uniqueness of our study is that it allows for a comprehensive comparison across countries and contexts and that the focus is not on beach recreation only, but instead also captures the marine environment more generally and includes both users and non-users. Although our welfare estimates appear low at first glance, these are figures for marine litter removal only, without taking into account preferences for achieving the GES of the other ten descriptors of the MSFD. However, estimates for the 11 descriptors cannot simply be added up, as achieving individual GES targets partly complement or hinder each other. Future studies should investigate how people trade-off the individual goals and what they would be willing to pay to achieve the GES of all eleven descriptors.

# Appendix

*Table A1: Attributes and attribute levels*

| Attribute | Description | Attribute levels |
|---|---|---|
| **Environmental status** | Plastic in the sea or ocean can harm both animals and plants. This affects the environmental status. The environmental status in 2030 can range between poor and excellent. | 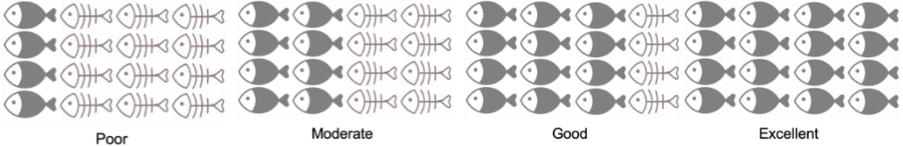 Poor — Moderate — Good — Excellent |
| **Certainty** | Reaching the desired environmental improvement through litter reduction by 2030 may not always be certain depending on the extent of the problem and the specific implemented measures. The certainty can range between 50%-99%. | 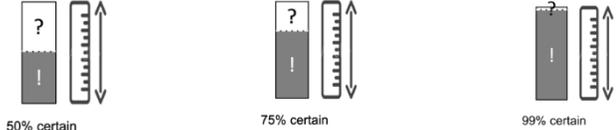 50% certain — 75% certain — 99% certain |
| **Size of plastic litter** | Measures can target different sizes of plastic litter. The focus can be on macro plastics or micro plastics or both. | 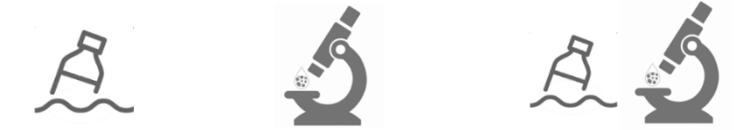 Remove macro plastics — Remove micro plastics — Remove both macro & micro plastics |
| **Geographic coverage** | Marine plastic removal can be addressed either in all European seas or in **[insert the country name where the survey is conducted]** sea alone. | 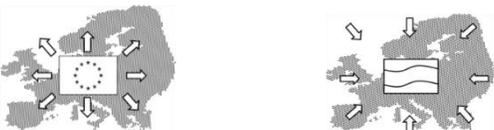 All European Seas — Your Country's Seas |

| Attribute | Description | Attribute levels |
|---|---|---|
| **Waste management measure** | In addition to cleaning up plastic litter from the sea, policies can prevent waste from ending up in marine environments through bans, improved collection (like deposit refunds) or better product labeling. | 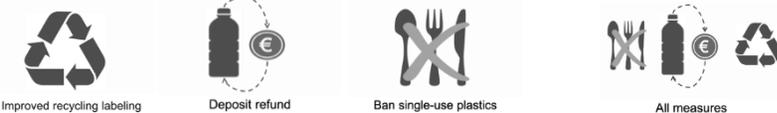 Improved recycling labeling — Deposit refund — Ban single-use plastics — All measures |
| **Additional waste management fee to your household** | The improvement will come at an additional cost to your household. Starting in 2021 your household fee for solid waste management will increase. The additional fee is to be paid for the next 10 years. Depending on the programm of measures, the cost range between € 5 and € 50 per month. | 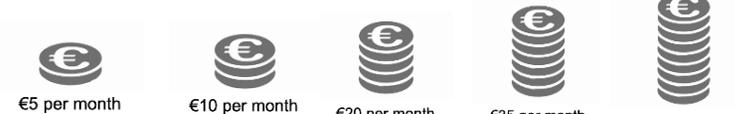 €5 per month (€60 per year) — €10 per month (€120 per year) — €20 per month (€240 per year) — €35 per month (€420 per year) — €50 per month (€600 per year) |

**Table A2: Estimates from mixed logit choice model across the 8 European countries with interaction terms of socio-demographic variables**

| | Denmark | Sweden | Estonia | Germany | The Netherlands | France | Italy | Greece |
|---|---|---|---|---|---|---|---|---|
| **CHOICE ATTRIBUTES** | Coef. (Std.err.) | Coef. (Std.err.) | Coef. (Std.err.) | Coef. (Std.err.) | Coef. (Std.err.) | Coef. (Std.err.) | Coef. (Std.err.) | Coef. (Std.err.) |
| **ASC$_{SQ}$** | -3.9682*** (0.1815) | -3.4916*** (0.1551) | -3.3217*** (0.1362) | -2.9309*** (0.1297) | -2.5972*** (0.1050) | -2.9329*** (0.1152) | -3.0236*** (0.1217) | -3.4728*** (0.1231) |
| **ENVIRONMENTAL STATUS (BASELINE CATEGORY: MODERATE)** | | | | | | | | |
| GOOD | 0.5519*** (0.1902) | 0.6221** (0.2831) | 0.2074 (0.1641) | 0.2534 (0.1687) | 0.3369** (0.1517) | 0.1062 (0.118) | 0.4745*** (0.1102) | 0.0515 (0.1180) |
| EXCELLENT | 0.8544*** (0.1994) | 1.2734*** (0.3375) | 0.5809*** (0.1738) | 0.9318*** (0.1872) | 0.6099*** (0.1729) | 0.2604* (0.1376) | 0.7057*** (0.1262) | 0.2649** (0.1286) |
| **CERTAINTY LEVEL (BASELINE CATEGORY: 50%)** | | | | | | | | |
| 75% | 0.2224** (0.0879) | 0.1936** (0.0904) | 0.0861 (0.0929) | 0.0594 (0.0871) | 0.0228 (0.0732) | 0.0078 (0.0788) | 0.0406 (0.0810) | 0.0828 (0.0708) |
| 99% | 0.3839*** (0.0914) | 0.3986*** (0.0915) | 0.2552*** (0.0945) | 0.2350*** (0.0884) | 0.2092*** (0.0762) | 0.1516* (0.0789) | 0.3951*** (0.0857) | 0.2538*** (0.0712) |
| **PLASTIC SIZE (BASELINE CATEGORY: MACROPLASTIC)** | | | | | | | | |
| MICROPLASTIC | 0.0832 (0.0682) | -0.0821 (0.0706) | -0.0128 (0.0675) | -0.0080 (0.0669) | -0.0606 (0.0564) | -0.0504 (0.0609) | 0.0797 (0.0635) | -0.1630*** (0.0573) |
| BOTH PLASTIC SIZES | 0.4398*** (0.0708) | 0.5208*** (0.072) | 0.2429*** (0.0695) | 0.3403*** (0.0692) | 0.1953*** (0.0563) | 0.2138*** (0.0583) | 0.4389*** (0.0667) | 0.0319 (0.0541) |
| **GEOGRAPHIC LOCATION (BASELINE CATEGORY: OWN COUNTRY)** | | | | | | | | |
| EU | 0.9837*** (0.1716) | 1.1205*** (0.1679) | 0.8386*** (0.1619) | 0.6369*** (0.1668) | 0.0329 (0.1279) | 0.1648 (0.1422) | 0.2635* (0.1573) | 0.4754*** (0.1310) |
| **WASTE MANAGEMENT MEASURE (BASELINE CATEGORY: ALL MEASURES)** | | | | | | | | |
| LABEL | -0.2108*** (0.0799) | -0.3856*** (0.0836) | -0.1500* (0.0819) | -0.2523*** (0.0805) | -0.0572 (0.0665) | -0.0924 (0.0709) | -0.1056 (0.0759) | -0.1797*** (0.0668) |
| REFUND | -0.1652** (0.0849) | -0.2768 (0.0843) | -0.1105 (0.0811) | -0.0829 (0.0806) | -0.0243 (0.0698) | -0.0351 (0.0731) | -0.0405 (0.0754) | -0.0318 (0.0683) |
| BAN | -0.2807*** (0.0865) | -0.3428*** (0.0929) | -0.2489*** (0.0867) | -0.1745** (0.0836) | -0.2133*** (0.0712) | -0.2229*** (0.0743) | -0.2297*** (0.0787) | -0.1000 (0.0690) |
| COST | -3.0434*** (0.1119) | -3.0877*** (0.1152) | -2.7311*** (0.0957) | -2.7552*** (0.099) | -3.2258*** (0.1064) | -3.1915*** (0.1231) | -3.452*** (0.1345) | -3.3520*** (0.1029) |

| (CONT.) | Denmark | Sweden | Estonia | Germany | The Netherlands | France | Italy | Greece |
|---|---|---|---|---|---|---|---|---|
| | Coef. (Std.err.) | Coef. (Std.err.) | Coef. (Std.err.) | Coef. (Std.err.) | Coef. (Std.err.) | Coef. (Std.err.) | Coef. (Std.err.) | Coef. (Std.err.) |
| **INTERACTION TERMS** | | | | | | | | |
| **COST*INCOME** | -1.70E-06** | -1.96E-06** | -2.03E-06 | -4.73E-06*** | -271E-06*** | -1.62E-06 | 6.49E-07 | -2.64E-06*** |
| | (8.22E-07) | (8.69E-07) | (1.76E-06) | (1.22E-06) | (7.31E-07) | (1.10E-06) | (1.24E-06) | (9.20E-07) |
| **EU*AGE** | -0.0046 | -0.0079** | -0.0078** | -0.0008 | 0.0077*** | 0.0024 | 0.0014 | -0.0058* |
| | (0.0032) | (0.0031) | (0.0031) | (0.0032) | (0.0026) | (0.0028) | (0.0031) | (0.0029) |
| **75%*FEMALE** | -0.0034 | 0.2082 | 0.2159* | 0.1906 | -0.0177 | 0.1283 | 0.2083* | 0.0860 |
| | (0.1224) | (0.1263) | (0.1219) | (0.1186) | (0.1026) | (0.1067) | (0.1114) | (0.1019) |
| **99%*FEMALE** | -0.0262 | 0.1911 | 0.0331 | 0.1825 | 0.0332 | 0.2024* | -0.0760 | 0.0123 |
| | (0.1261) | (0.1264) | (0.1224) | (0.1195) | (0.1058) | (0.1073) | (0.1147) | (0.1015) |
| **VISIT *GOOD** | 0.2895 | 0.2477 | 0.2351 | 0.3752** | -0.0229 | 0.2676** | -0.1298 | 0.0515** |
| | (0.2011) | (0.2906) | (0.1764) | (0.1810) | (0.1603) | (0.1319) | (0.1265) | (0.1180) |
| **VISIT *EXCELLENT** | 0.3832* | 0.0676 | 0.2240 | 0.2926 | 0.1033 | 0.5887*** | 0.0775 | 0.2649** |
| | (0.2083) | (0.3428) | (0.1856) | (0.1973) | (0.1826) | (0.1549) | (0.1419) | (0.1286) |
| **SD** | | | | | | | | |
| **ENVIRONMENTAL STATUS (BASELINE CATEGORY: MODERATE)** | | | | | | | | |
| **EXCELLENT** | 0.4183** | 0.9151*** | 0.6524*** | -0.617*** | 0.7734*** | -0.6972*** | 0.7275*** | -0.0595*** |
| | (0.1498) | (0.1848) | (0.1875) | (0.2834) | (0.1744) | (0.1612) | (0.1577) | (0.2648) |
| **GOOD** | -0.0089 | -0.0025 | 0.324* | 0.0243 | 0.0212 | -0.0073 | 0.0101 | 0.5930 |
| | (0.1636) | (0.1193) | (0.1372) | (0.1329) | (0.1024) | (0.1117) | (0.1148) | (0.1071) |
| **CERTAINTY LEVEL (BASELINE CATEGORY: 50%)** | | | | | | | | |
| **75%** | 0.0109 | 0.0142 | -0.0062 | 0.0166 | 0.0019 | 0.0211 | -0.0074 | -0.0026 |
| | (0.1396) | (0.1175) | (0.1236) | (0.1379) | (0.1145) | (0.1171) | (0.1415) | (0.0974) |
| **99%** | 0.0388 | -0.0359 | 0.1509 | 0.0393 | -0.3625** | -0.1462 | -0.3939** | -0.0125 |
| | (0.2042) | (0.195) | (0.1727) | (0.2064) | (0.1448) | (0.2932) | (0.1581) | (0.1400) |
| **PLASTIC SIZE (BASELINE CATEGORY: MACROPLASTIC** | | | | | | | | |
| **MICROPLASTIC** | 0.5979*** | 0.7314*** | 0.638*** | -0.6295*** | -0.3388** | 0.6323*** | 0.7389*** | 0.5726*** |
| | (0.1389) | (0.1319) | (0.1355) | (0.1317) | (0.1717) | (0.1136) | (0.1136) | (0.1136) |
| **BOTH PLASTIC SIZES** | 0.4082** | 0.1267 | -0.477** | -0.5644*** | -0.0679 | -0.0445 | -0.5464*** | 0.0053 |
| | (0.1999) | (0.3827) | (0.1861) | (0.1614) | (0.2765) | (0.529) | (0.1438) | (0.1759) |
| **GEOGRAPHIC LOCATION (BASELINE CATEGORY: OWN COUNTRY)** | | | | | | | | |
| **EU** | 0.8506*** | 0.8322*** | 0.7059*** | 0.8141*** | 0.6508*** | -0.6683*** | 0.7159*** | 0.5133*** |
| | (0.0902) | (0.0945) | (0.095) | (0.0915) | (0.0786) | (0.0823) | (0.085) | (0.0834) |

| (CONT.) | Denmark | Sweden | Estonia | Germany | The Netherlands | France | Italy | Greece |
|---|---|---|---|---|---|---|---|---|
| | Coef. (Std.err.) | Coef. (Std.err.) | Coef. (Std.err.) | Coef. (Std.err.) | Coef. (Std.err.) | Coef. (Std.err.) | Coef. (Std.err.) | Coef. (Std.err.) |
| **WASTE MANAGEMENT MEASURE (BASELINE CATEGORY: ALL)** | | | | | | | | |
| LABEL | -0.7028*** (0.1633) | 0.8569*** (0.1563) | 0.8706*** (0.1623) | -0.8279*** (0.1614) | -0.3816* (0.221) | -0.6769*** (0.1529) | -0.8509*** (0.1491) | 0.5894*** (0.1514) |
| REFUND | 0.0119 (0.2031) | -0.0461 (0.2164) | 0.0769 (0.2135) | -0.0143 (0.1940) | 0.0019 (0.1839) | 0.0059 (0.1715) | 0.0189 (0.1579) | -0.0010 (0.1541) |
| BAN | -0.1559 (0.2709) | 0.9548*** (0.1485) | 0.6561*** (0.1685) | 0.2465 (0.3193) | 0.3039 (0.2263) | -0.3537* (0.2011) | 0.5825*** (0.1654) | 0.2774 (0.2504) |
| COST | 1.2680*** (0.0767) | 1.2691*** (0.0726) | 1.2989*** (0.0681) | 1.2906*** (0.0722) | 1.4774*** (0.0823) | 1.5875*** (0.0937) | 1.5585*** (0.0985) | 1.6867*** (0.0842) |
| NUMBER OF OBSERVATIONS | 14,994 | 15,984 | 16,254 | 16,254 | 16,524 | 16,470 | 16,740 | 17,028 |
| NUMBER OF RESPONDENTS | 833 | 888 | 903 | 903 | 918 | 915 | 930 | 946 |
| LR CHI2(10) | 941.97 | 955.27 | 1380.59 | 1431.34 | 1267.82 | 1498.34 | 1147.39 | 1588.79 |
| PROB>CHI2 | 0.000 | 0.000 | 0.000 | 0.000 | 0.000 | 0.000 | 0.000 | 0.000 |
| LOG LIKELIHOOD | -2986.264 | -3371.660 | -3755.814 | -3724.658 | -4214.601 | -4148.800 | -4057.607 | -4096.421 |
| AIC | 6030.527 | 6801.320 | 7569.628 | 7507.315 | 8487.201 | 8355.600 | 8173.215 | 8250.842 |
| BIC | 6251.374 | 7024.021 | 7792.815 | 7730.502 | 8710.866 | 8579.641 | 8397.256 | 8475.378 |

*Note: In the MXL model all attributes were taken as random parameters; The number of observations is lower due to missing information on some of the socio-economic variables;\*= statistically significant at the 10% level; \*\*= 5% level; and \*\*\* = 1% level.*

**Table A3: Marginal WTP**

| | Denmark | Sweden | Estonia | Germany | The Netherlands | France | Italy | Greece |
|---|---|---|---|---|---|---|---|---|
| | Coef. (Std.err.) (Conf.int 95%) | Coef. (Std.err.) (Conf.int 95%) | Coef. (Std.err.) (Conf.int 95%) | Coef. (Std.err.) (Conf.int 95%) | Coef. (Std.err.) (Conf.int 95%) | Coef. (Std.err.) (Conf.int 95%) | Coef. (Std.err.) (Conf.int 95%) | Coef. (Std.err.) (Conf.int 95%) |
| **ENVIRONMENTAL STATUS (BASELINE CATEGORY: MOERATE)** | | | | | | | | |
| GOOD | 0.0815*** (0.0101) (0.0618 - 0.1012) | 0.0778*** (0.0105) (0.0573 - 0.0983) | 0.0605*** (0.0111) (0.0387 - 0.0822) | 0.0856*** (0.0125) (0.0612 - 0.1100) | 0.0411*** (0.0081) (0.0252 - 0.0570) | 0.0456*** (0.0098) (0.0265 - 0.0648) | 0.0396*** (0.0074) (0.0250 - 0.0541) | 0.0386*** (0.0092) (0.0204 - 0.0568) |
| EXCELLENT | 0.1183*** (0.0134) (0.0919 - 0.1446) | 0.1222*** (0.0151) (0.0925 - 0.1518) | 0.1148*** (0.0156) (0.0842 - 0.1454) | 0.1687*** (0.0202) (0.1291 - 0.2084) | 0.0861*** (0.0126) (0.0614 - 0.1109) | 0.1071*** (0.0158) (0.0760 - 0.1382) | 0.0834*** (0.0112) (0.0615 - 0.1053) | 0.0839*** (0.0136) (0.0572 - 0.1105) |
| **CERTAINTY LEVEL (BASELINE CATEGORY: 50%)** | | | | | | | | |
| 75% | 0.0226*** (0.0064) (0.0101 - 0.035) | 0.0257*** (0.0066) (0.0127 - 0.0387) | 0.0295*** (0.0098) (0.0103 - 0.0488) | 0.0225** (0.009) (0.0048 - 0.0402) | 0.0016 (0.0066) (-0.0113 - 0.0145) | 0.0102 (0.0082) (-0.0059 - 0.0264) | 0.0146** (0.0064) (0.0021 - 0.0271) | 0.0178** (0.0082) (0.0017 - 0.0388) |
| 99% | 0.0377*** (0.0071) (0.0237 - 0.0517) | 0.0473*** (0.0079) (0.0316 - 0.063) | 0.0388*** (0.0099) (0.0192 - 0.0583) | 0.0461*** (0.099) (0.0266 - 0.0657) | 0.0307*** (0.0078) (0.0155 - 0.0459) | 0.0397*** (0.0095) (0.0211 - 0.0583) | 0.0362*** (0.0074) (0.0217 - 0.0507) | 0.0394*** (0.0092) (0.0212 - 0.0576) |
| **PLASTIC SIZE (BASELINE CATEGORY: MACROPLASTIC)** | | | | | | | | |
| MICROPLASTIC | 0.0074 (0.0063) (-0.0049 - 0.0198) | -0.0080 (0.0064) (-0.0205 - 0.0045) | -0.0003 (0.0097) (-0.0193 - 0.0187) | -0.0015 (0.0092) (-0.0195 - 0.0164) | -0.0076 (0.0069) (-0.0211 - 0.0060) | -0.0072 (0.0089) (-0.0246 - 0.0102) | 0.0085 (0.0067) (-0.0046 - 0.0215) | -0.0242** (0.0090) (-0.04186 - -0.0065) |
| BOTH PLASTIC SIZES | 0.0445*** (0.0075) (0.0297 - 0.0593) | 0.0464*** (0.0079) (0.0309 - 0.0619) | 0.0329*** (0.0105) (0.0124 - 0.0536) | 0.0500*** (0.0108) (0.0289 - 0.0711) | 0.0256*** (0.0074) (0.0112 - 0.0401) | 0.0306*** (0.0091) (0.0127 - 0.0484) | 0.0476*** (0.0083) (0.0313 - 0.0639) | 0.0043 (0.0080) (-0.0114 - 0.0201) |
| (CONT.) | Denmark | Sweden | Estonia | Germany | The Netherlands | France | Italy | Greece |

| | Coef. (Std.err.) (Conf.int 95%) | Coef. (Std.err.) (Conf.int 95%) | Coef. (Std.err.) (Conf.int 95%) | Coef. (Std.err.) (Conf.int 95%) | Coef. (Std.err.) (Conf.int 95%) | Coef. (Std.err.) (Conf.int 95%) | Coef. (Std.err.) (Conf.int 95%) | Coef. (Std.err.) (Conf.int 95%) |
|---|---|---|---|---|---|---|---|---|
| **GEOGRAPHIC LOCATION (BASELINE CATEGORY: OWN COUNTRY)** | | | | | | | | |
| **EU** | 0.0709*** (0.0087) (0.0539 - 0.0879) | 0.0685*** (0.0091) (0.0507 - 0.0863) | 0.0690*** (0.0107) (0.0481 - 0.0899) | 0.0833*** (0.0117) (0.0604 - 0.1062) | 0.0478*** (0.0081) (0.0318 - 0.0637) | 0.0413*** (0.0087) (0.0241 - 0.0584) | 0.0336*** (0.0064) (0.0210 - 0.0462) | 0.0356*** (0.0078) (0.0201 - 0.0510) |

*= statistically significant at the 10% level; **= 5% level; and *** = 1% level